\documentclass[
 reprint,
secnumarabic,
 amssymb, amsmath,
 aip,rsi
groupedaddress,
frontmatterverbose,
]{revtex4-1}
\usepackage{graphicx}
\usepackage{bm}%
\usepackage[version=4]{mhchem}
\usepackage[colorlinks=true,linkcolor=blue]{hyperref}%
\expandafter\ifx\csname package@font\endcsname\relax\else
 \expandafter\expandafter
 \expandafter\usepackage
 \expandafter\expandafter
 \expandafter{\csname package@font\endcsname}%
\fi
\begin{document}

\title{A new method for measurement and quantification of tracer diffusion in  nanoconfined liquids }%

\author{V. J. Ajith}%
\affiliation{Department of Physics, Indian Institute of Science Education and Research, Pune 411008, Maharashtra, India}%
\author{Shivprasad Patil}
\email {s.patil@iiserpune.ac.in}
\affiliation{Department of Physics, Indian Institute of Science Education and Research, Pune 411008, Maharashtra, India}

\begin{abstract}
  We report development of a novel instrument to measure tracer diffusion in water under nano-scale confinement.  A direct optical access to  the confinement region, where water is confined between a tapered fiber and a flat substrate, is made possible by coating the probe with metal and opening a small aperture( 0.1 $\mu$m to 1 $\mu$m) at its end. A well-controlled cut using an ion beam ensures desired lateral confinement area as well as adequate illumination  of the confinement gap. The probe is mounted on a tuning-fork based force sensor to control the separation between the probe and the substrate with nm precision.  Fluctuations in fluorescence intensity due to diffusion of a dye molecule in water confined between probe and the sample are recorded using a confocal arrangement with a single photon precision.  A Monte-Carlo method is developed to determine the diffusion coefficient from the measured autocorrelation of intensity fluctuations which accommodates the specific geometry of confinement and the illumination profile. The instrument allows measurement of diffusion laws under confinement. We found that the diffusion of a tracer molecule is slowed down by more than ten times for the probe-substrate separations of 5 nm and below.

\end{abstract}
\maketitle
\section{Introduction}
 Physical properties of  liquids under nanoscale confinement are debated in the literature. The behaviour of  nano-confined liquids is  relevant in  areas ranging from tribology to biology \cite{Finney, Isra1, Persson, Granick}. The confinement by solid walls is reported to cause significant change in liquid's physical properties such as diffusion coefficient, viscosity and dielectric constant \cite{Ashish,Granick2, Uzi}, and it undergoes a phase transition leading to  solidification at room temperature \cite{Granick2, Klein}. However, both the phenomenon and its cause  is debated for decades \cite{Isra-Horn,Isra-Christ,Grani-phys}.  In particular, the flow properties under confinement using different experimental approaches have led to contradictory  results \cite{Derjaguin, Isra2, Granick3, Raviv1, Raviv2}.  The Surface Force Apparatus(SFA) measurements  have reported  that  water shows no change in viscosity for film thickness as small as  $0.4$ nm and below\cite{Raviv1, Raviv2}. However, many other measurements  have reported orders of magnitude increase in viscosity under confinement\cite{Granick2,Uzi}.Nano-confined thin films of non-associative fluids such as Octamethylcyclotetrasiloxane(OMCTS) exhibit significant non-linear effects in their viscosity measurements\cite{Granick2}. The small-amplitude AFM measurements of both water and OMCTS layers under confinement have revealed a speed dependant dynamic solidification\cite{Patil1,Khan}. The lipidic cubic phases provides a versatile porous media to investigate the transport of molecules through the confined water\cite{mazzenga_exp1,mazzenga_exp2,mazzenga_sim}.More recently, it is argued that contradiction among different experimental reports  stem from working with different experimental parameters such as  shear frequency, shear rates and  wettability of confining walls\cite{Grani-phys,aman_jpcm}.
 
 To gain insight into the mechanisms driving the change in flow properties under confinement, it is also important to develop understanding of molecular mobility under confinement. There are many SFA or AFM-like techniques to measure the viscous stress in confinied liquids upon straining the liquid film. However, there are not many tools for measurement of  translational or rotational diffusion under confinement.

 In literature there are scant reports of direct measurement of tracer diffusion under nano-confinement. The nanochannels are used to confine liquid and  optical methods  have been employed to measure diffusion in $\sim$ 10 nm confinement\cite{Santo, Zhong}. Simulations are performed to understand the transport of solute molecules through periodic array of nanopores formed by inverse bicontinuous lipidic  cubic phases.\cite{mazzenga_sim} Experimentally  the diffusion of solute molecules through them  is measured and was found to depend on solute molecular structure and confinement geometry\cite{mazzenga_exp1,mazzenga_exp2}.Molecular dynamics simulations are performed to understand the diffusion of water between two graphene sheets separated by 1 nm and in silica nanopores\cite{graphene_slit,silica_nanopore}.      In diffusion measurement using  SFA  the gap formed between two surfaces mounted on piezoelectric transducers is used for confinement\cite{Ashish} In AFM-like confining geometry,  the lateral confinement size is much smaller than the  diffraction limit and hence it is difficult to access this volume optically for diffusion measurement. On the other hand active feedback control can be used to hold the tip and substrate at sub 10 nm separations -a challenging task owing to the soft springs used to measure normal surface forces  in case of SFA.  The trade-off between sensitivity and stability makes it difficult to employ servo control to hold the separation between surfaces at a desired value.  We need an  instrument that can maintain the separation between surfaces below 5 nm with an active control and having an optical access to the probe-substrate gap.    
 
 This paper reports development of a new instrument with near-infinite normal stiffness to avoid the possible jump-in instability and ability to measure tracer diffusion of fluorescent dyes in water confined to less than 10 nm.   We  use  tuning fork based  force sensor to measure the shear response of the nanoconfined water. A single mode optical fiber is  tapered at one end to form a tip and coated with Aluminum. A small ($\leq$ 1 $\mu$m) optical aperture  is opened at the tapered end using Focused Ion Beam(FIB). This tip with an optical aperture  forms one of the confining surfaces and it is roughly 10 times bigger in lateral size compared to AFM or Near-field Scanning Optical Microscope (NSOM) probes. It allows the gap to be illuminated with significantly larger power and the illumination is far field  as opposed to NSOM probes which are designed for better lateral resolution to beat the diffraction limit.   The other surface is optically flat and transparent substrate.  The well-controlled opening of an aperture using FIB  ensures that the confinement area is larger than the optically illuminated area.  The tip is mounted on one prong of the tuning fork.  The other prong is driven mechanically using a piezo-drive.   We add a small amount of a tracer dye ( 100 nM- 1 $\mu$M ) into the water to be confined between two surfaces.  Using a confocal  arrangement,  the fluorescence from the molecule diffusing in the tip-substrate gap is recorded using Avalanche Photo-Diode (APD). The fluctuations in fluorescence due to diffusion of the molecule/s  are recorded and used to calculate the autocorrelation.  The instrument allows measurement of diffusion under nano-confinement with a lateral size much larger than in case of AFM probes. The confinement area approaches close to  a typical SFA measurement($0.1$ to $1$ $\mu$m). Furthermore, a Monte-Carlo simulation method is developed to obtain diffusion coefficient from the experimental data. The simulation takes into account the confinement geometry and its parameters as well as the illumination profile specific to our experiments. Few initial results are presented in which we observe a significant slow-down in mobility  below confinement of 5 nm.     
 
 The paper is organised in the following manner. We first discuss the design and development of the instrument to measure diffusion in nano-confined water. Next, We describe the Monte-Carlo method developed for data analysis. We present initial results on tracer diffusion in water which is confined in a gap of less than 5 nm using the instrument. The results are discussed in the context of similar efforts in the literature.

\section{THE INSTRUMENT}
The instrument consists mainly of three parts. i) A fiber tip prepared from a single mode optical fiber to illuminate nano-confined water, ii) a feedback mechanism and servo control  and iii) the necessary illumination and collection optics to collect the light from confined region.           

\subsection{The  fiber tip }
\begin{figure}
      \includegraphics[width=0.4\textwidth]{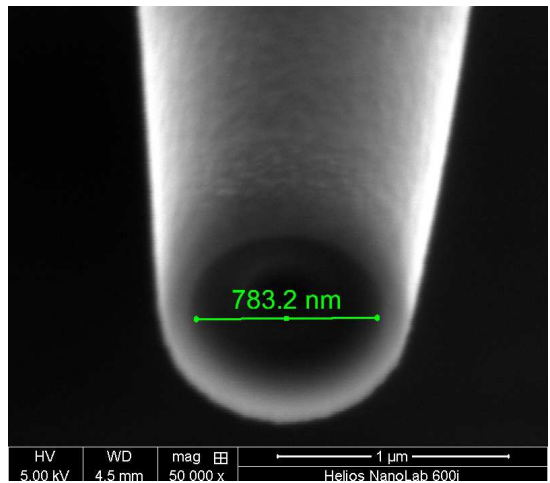}
  \caption{Scanning Electron Microscopy image of a typical optical fiber probe  used in the instrument. A metal coated tapered single mode optical fiber is sliced at the end to open an optical aperture of $\sim$ 800nm   \label{sem}}
\end{figure}

The optical fiber probe is made out of a single mode optical fiber (Thorlabs 460HP, diameter without buffer=$125 \mu$m).  A sharp tip is prepared  by pulling a fiber using a laser based fiber puller (Sutter P2000). The process is optimized to give conical tips with $40$ nm diameter. The tips are coated with a layer of $\sim$150nm aluminum using sputter coating method. The end of coated tips are then sliced using FIB to open up an optical aperture. We get a flat surface at the end of the tip with an optical aperture in the centre with a diameter of $\sim$ 100nm to 1 $\mu m$. Figure \ref{sem} shows SEM picture of a typical optical fiber probe. The radius of the optical aperture depends on the position at which we slice the fully coated tip. The data shown in this paper are obtained with a tip having 800 nm as the aperture diameter, prepared in this way. 

A fiber with a tip at its one end  is mounted on the free prong of the tuning fork. The tuning fork is fixed to the scanner piezo tube.  The other end of the fiber  is passed through the glass tube and scanner piezo. It is further connected to a a single-mode fiber using a splicer.  It  serves the purpose of both taking the light from the gap into the photo-diode or carrying the laser light into the tip-sample gap depending on the configuration. 

\subsection{Feedback}

A feedback system is necessary to control the gap between the optical fiber probe and the substrate where the water is confined. For implementing such a feedback, we need mechanisms for (1) Sense a signal proportional to separation and (2)adjust the tip-substrate separation. 
\subsubsection{Inertial sliding for nano-actuation } 
\begin{figure}
      \includegraphics[width=0.5\textwidth]{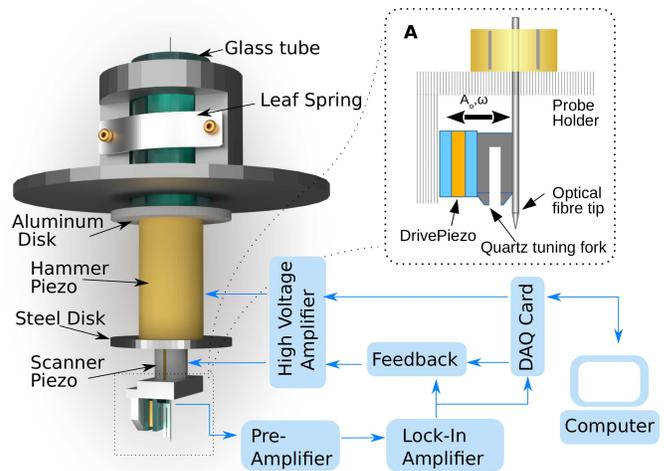}
  \caption{ The schematic of piezo-assembly used for nano-actuation and the implementation of the feedback. The inset shows the schematic of tuning fork sensor. The current through prongs is used as process variable in the feedback.\label{schematic}}
\end{figure}

The relative separation between the tip and cover-slip can be adjusted either by moving the tip or the substrate.  We reported an instrument to measure rheological response of nanoconfined liquids wherein the force sensor was above the substrate and fine and coarse approach using piezo tubes were situated beneath it.\cite{Karan1}.  This arrangement is not suitable for optical measurement as it blocks the optical access to tip-sample gap from both top and bottom.  We have now modified the set-up so that both, the force sensor (tuning fork bearing the tip) and actuation assembly,  are above the substrate.  We integrated the piezo tube assembly and tuning fork sensor as shown in the figure \ref{schematic}. The new design retains all the advantages of inertial sliders for coarse and fine approach and frees up the space below the substrate for placing optical components such as objective lens and pinholes. In the new design, the force sensor bearing the tip moves during the  measurement  and  substrate remains fixed.

 Figure \ref{schematic} shows the inertial actuation assembly based on principle of inertial sliding. It consists of following parts. There are two concentric piezo tubes mounted on an Aluminum disk. The inside piezo is a typical quadrant tube for the purpose of X-Y scanning and Z fine movement. It is called scanner piezo. The outer tube functions as a hammer in the coarse movement. The upper end of this disc is connected to  quartz glass tube which slides through stainless steel tube-holder. The glass tube is held onto the holder using a leaf-spring of berillium-copper alloy. The screws on the leaf spring are used to adjust tension so that the tube along with piezo assembly is held up from sliding through the holder under gravity.  The coarse movement along z-axis is achieved by  an inertial sliding mechanism. For approach(retract) , the voltage is slowly increased (decreased) from zero to expand(contract) the hammer piezo. The sharp drop of voltage to zero, results  in a quick contraction (expansion) of piezo, sliding the glass tube downwards(upwards) while retaining the position of steel disk due to inertia. This causes the  entire assembly to move down (up) in each pulse. Thus the tip is brought towards(away from) the surface. By playing with leaf-spring tightness, pulse height, pulse length and number of pulses in a burst we can control the movement in each step with precision  of few hundred nm. The entire range of the motion is few mm.

The fine movement of the tip while performing the experiments is achieved using the scanner piezo. Its sensitivity is measured with a fiber based interferometer and it is used to calculate the separation between the tip and substrate. The details of this procedure can be found elsewhere.\cite{pre_karan}

The entire assembly for actuation is mounted on a opto-mechanical cage plate. This cage plate can be moved in X,Y,Z directions  using micrometer screw gauges having  precision  of 0.01 mm.  Using a combination of pulses to both hammer and scanner piezo, the tip is approached towards the surface from distance of few mm to within a nm. 

\subsubsection{Sensor} 
The optical fiber probe is attached to one prong of tuning fork and the other prong is attached to a dither piezo  as shown in the inset of figure \ref{schematic}. Using this piezo the tuning fork with fiber is oscillated laterally at its resonance frequency. The resonant oscillations results in a  $90^o$ out of phase bending between the prongs. This produces an oscillating current in the tuning fork, whose amplitude and phase is measured using a Lock-In amplifier(Signal Recovery 7265). With the help of an actuation assembly, the oscillating tip is brought close to the substrate. As the tip is approached within 20 to 30 nm from the substrate, the oscillating amplitude of tip-bearing prong decreases, which reduces the current. This monotonic decrease in  current amplitude with respect to separation is used as input signal to  feedback control. This allows holding the tip at a fixed separation while measurements are performed. Feedback is implemented using an  home built op-amp based PI feedback circuit. The program to interface the instrument to electronics  via a DAQ card(NI PXI-6259) is written in LabVIEW.   A  fine and coarse approach program is implemented to auto-approach the tip to a desired set-point  amplitude, which determines the probe-substrate separation in the range of 1-20 nm.

\subsection{The optics }
\begin{figure}
      \includegraphics[width=.45\textwidth]{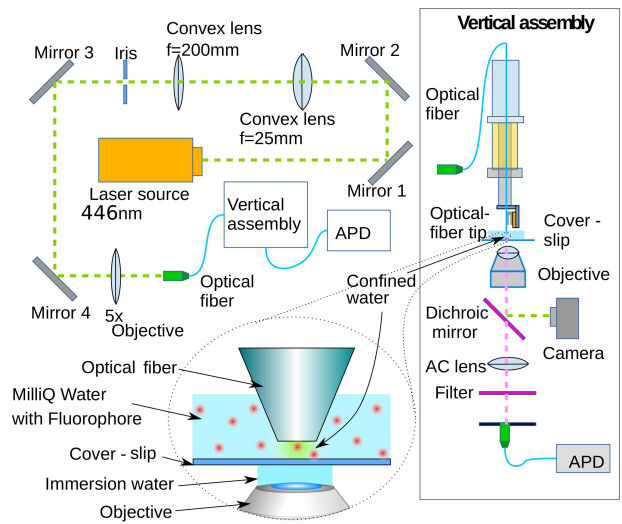}
  \caption{The schematic of the optics used for collection of florescence from the tip-sample junction. The illumination is performed by guiding the laser light into the tip and the fluorescence is collected by an objective from below. An APD is used to record the fluorescence intensity fluctuations.\label{optics}}
\end{figure}

The optical setup is illustrated schematically in the figure \ref{optics}. The liquid is confined between the tip and the substrate. The diameter of the lateral circular area of confinement is about 1 $\mu$m. Part of this confined liquid  is illuminated using the optical aperture of 100 - 800 nm optical opening at the tip.   The other end of  the tip-bearing fiber is  cleaved and mechanically spliced to  single mode fiber (CORELINK Fiber Optic Mechanical Splice). It is then coupled to pass laser light (wavelength 446 nm or 532 nm, power 15mW). The coupling of the fiber to the laser  is accomplished  by focusing  a collimated beam onto the end of this fiber. See figure \ref{optics}  for details. The mechanically spliced arrangement allows the change of tip without disturbing the laser alignment. 

 For collection,  a  60x water immersion objective with a numerical aperture of 1.2 (Olympus 60X UPlanSApo) is kept below the cover-slip. It is used to collect light from the region where liquid is confined between the tip and cover-slip.  A dichroic is used to separate the fluorescence emission of tracer molecules  from  excitation light. The emitted light is further focused onto the core (25 $\mu$m) of a multimode fiber connected to Avalanche Photo Diode (Perkin Elmer SPCM-AQRH-13-FC). The  core  of the multimode  fiber acts as a pinhole and its diameter determines  confocal volume from where light is collected.   The photon counts from APD are then fed into a correlator card (Flex99OEM-12/D) to  compute the autocorrelation of intensity fluctuations. Autocorrelation curves  are then transferred and analyzed in a computer using a LabVIEW program. 

It is possible to invert the illumination path. Instead of illuminating the observation volume with optical fiber tip,  we can  do it using the objective. This configuration is similar to the conventional Fluorescence Correlation Spectroscopy(FCS). This is accomplished  by redirecting the expanded parallel beam to reflect from the dichroic filter to the objective. Thus objective can illuminate same focal volume selected by the confocal arrangement. This can be done by introducing few extra mirrors without disturbing the original alignment in the optical setup. This helps in measuring the diffusion coefficient of probe molecule in bulk water.

\section{MEASUREMENTS} 
\subsection{Experimental data} 

\begin{figure}
      \includegraphics[width=.45\textwidth]{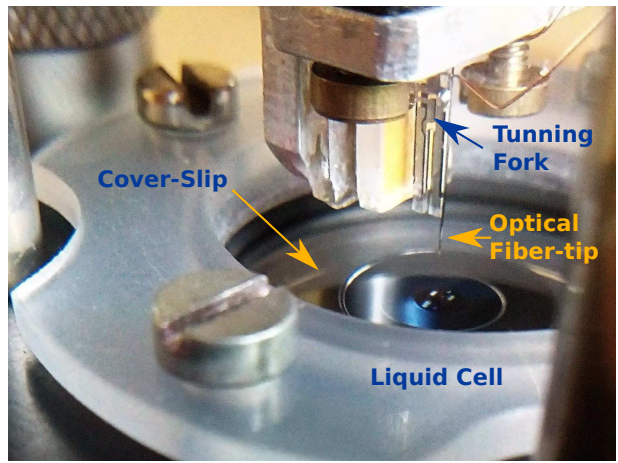}
  \caption{A photograph of tuning fork and the tip assembly along with the liquid cell. \label{liquidcell}}
\end{figure}
 
Figure \ref{liquidcell} shows an image of tuning fork sensor and the liquid cell assembly. The cover slips used in the experiment are characterized using AFM for estimating their roughness. The rms roughness is 0.5$\pm$0.2 nm 
It is used as the bottom confining surface and it is fitted into liquid cell.  It consists of a slotted aluminum with a hole in the centre which facilitates collection of light from below. The cover-slip is held in place by a pair of pressed viton o-rings from its both sides. It is pressed by the aluminum plate from below and a Kel-F sheet with opening in it from the top. The opening serves the purpose of lowering the tuning fork assembly into the cell and allowing the contact between the tip and the substrate(cover-slip). The cover-slip is first cleaned with methanol and then kept under UV cleaner for 30 minutes. It is then immediately  fixed into the cell and filled with $100\mu l$ solution of  100 nM tracer dye such as Rhodamine 6G (\ce{C28H31ClN2O3}) or Coumarin 343(\ce{C16H15NO4}) in  Milli-Q water. The hydrodynamic radii of these dye molecules are 0.6 nm \cite{rhod6g_radius}and 0.4 nm \cite{c343_radius} respectively. The objective is focused on the upper surface of the cover-slip with the help of camera. The fiber-tip  is immersed in the water and brought close to the cover-slip in a controlled manner so that the tuning fork prongs stay out of water.  The amplitude of the tip-bearing prong is recorded as the tip approaches the surface. The tip is kept at a constant distance from the substrate within few nm using the amplitude feedback.  The fluorescence from the tracer molecule/s  in confined water is recorded using an APD. 

\subsection{Autocorrelation of intensity fluctuations}
\begin{figure}
      \includegraphics[width=0.45\textwidth]{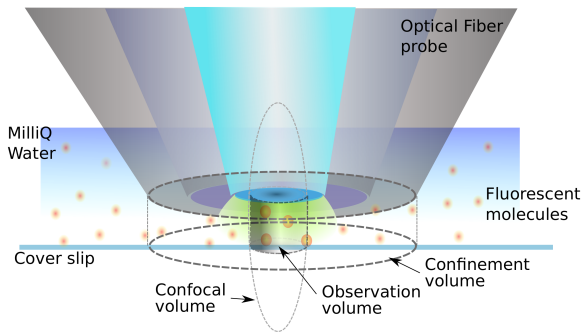}
  \caption{The schematic of the confinement gap formed by tip-substrate junction and tracer molecules diffusing through it. The observation volume is convolution of the confocal volume and the illumination profile. The height of this volume is determined by tip-substrate gap ($\leq$ 5 nm) and the lateral size by width of confocal volume. \label{confinement}}
\end{figure}

The  autocorrelation  of intensity  fluctuations over a period of 1s is calculated by a correlator card and is repeated over 20 times. The average of 20 correlation curves is taken.  The process is repeated for different separations, typically from 1 to 5 nm.  The separation is controlled by varying the set-point amplitude in feedback controller and determined by noting the z-voltage on the scanner piezo. Figure \ref{confinement} shows schematic of confinement and tracer molecules diffusing through it. 
Our instrument allows illumination of a small observation volume in confined water and detect the fluorescent photons from it, which is related to  number of tracer molecules in the observation volume. The tracer molecules diffuse in and out of the  observation volume resulting in fluctuation of fluorescence intensity measured using APD. Since these fluctuations are a result of the diffusing tracer molecules, the autocorrelation of fluorescence intensity can provide diffusion coefficient of tracer molecules under confinement, provided that the confinement geometry and size is experimentally determined. This method is similar to Fluorescence Correlation Spectroscopy (FCS) used for measurement of diffusion coefficient, reaction kinetics and microrheology.\cite{Magde1,Magde2,Magde3,FCSreview,FCSmicrorheo}. 

  In conventional  FCS, the tracer molecule can enter the observation volume from any direction. If the diffusion is fickian and the confocal observational volume approximated as a 3-D Gaussian of height of $\sim 1  \mu$m and width in lateral plane of  $ \sim  \lambda/2 $, then the autocorrelation of the fluorescent intensity fluctuations is derived as 

\begin{align}
G(\tau)&=\frac{1}{<N>}\left(1+\frac{\tau}{\tau_D}\right)^{-1}\left(1+r^{2}\frac{\tau}{\tau_D}\right)^{{-1}/{2}}\label{eq:3dfcs_td}\\
&=\frac{1}{<N>}\left(1+\frac{4\tau D}{\omega_{xy}^2}\right)^{-1}\left(1+\frac{4\tau D}{\omega_{z}^2}\right)^{{-1}/{2}}\label{eq:3dfcs_wxy}
\end{align}
 
where $\tau_D$, the diffusion time, is the average residence time of the tracer molecule in  the observation volume.  $\tau_D=\frac{\omega_{xy}^2}{4D}$, where $D$ is the diffusion coefficient and $\omega_{xy}$ is the radius in xy-plane and $\omega_z$ is height of the observation volume. $\omega_{xy}$ and $\omega_z$ are usually determined  by recording autocorrelations for tracer dye whose $D$ is known and fitting it to equation \ref{eq:3dfcs_wxy}.

Ignoring the experimental constraint on the tracer molecule that it cannot cross the physical boundary imposed by the tip and substrate, we can use \ref{eq:3dfcs_wxy} to estimate diffusion coefficient by fitting the experimental data. If the diffusion takes place only in a 2D plane then the equation \ref{eq:3dfcs_wxy} becomes. 
\begin{align}
G(\tau)&=\frac{1}{<N>}\left(1+\frac{4\tau D}{\omega_{xy}^2}\right)^{-1}\label{eq:2dfcs_wxy}
\end{align}

Since the illumination profile and the observation volume is different in our experiments, the validity of equation \ref{eq:3dfcs_wxy} or \ref{eq:2dfcs_wxy} to estimate diffusion coefficient by fitting it to our experimental data can be questioned. Secondly, the confinement in z-direction by physical boundary does not allow the fluorophore to freely escape or enter from this direction. The experimental situation in our case is different from the assumptions used to derive \ref{eq:3dfcs_wxy} or \ref{eq:2dfcs_wxy}, yet the approximate diffusion estimated from the fit of this  equation is also  meaningful. Although it is less accurate, this fitting method is much simpler to estimate the apparent diffusion coefficient.  
\section{SIMULATIONS}

Since  equation \ref{eq:3dfcs_wxy} is derived for a specific excitation profile and observation volume in which molecules freely diffuse in and out, fitting it to the autocorrelation curves obtained in our experiments may  not give correct diffusion coefficient.  In our setup the tracer diffusion in  z-direction is restricted by the confining boundaries. This restriction may cause multiple reflections from the boundary and it may spend more time in the observation volume leading to inaccurate diffusion coefficient. The residence time $\tau_D$ is determined by both diffusion and the restriction in z-motion. The other important factor is the excitation profile in our experiments. Equation \ref{eq:3dfcs_wxy} is derived for an objective illumination. In our setup the illumination is done using the fiber-tip and emission is collected using  a confocal arrangement. The actual observation volume will be a convolution of tip-illumination and the confocal volume. 

 For the above-mentioned  reasons it is necessary to develop a method to extract diffusion coefficient from experimental autocorrelation curves which will accommodate our excitation profile and impose the physical boundaries on top and bottom within which the tracer molecules are restricted.   We develop a Monte-Carlo method for analyzing our data which has flexibility to change the illumination profile and accommodates the effects of confining boundaries. The autocorrelation curves obtained from multiple simulation runs are fitted to experimental data to extract accurate diffusion coefficient from the measurements.

\begin{figure}
      \includegraphics[width=0.45\textwidth]{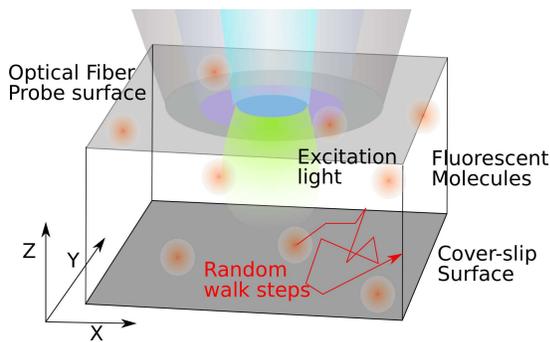}
  \caption{The simulation box. The fluorescing tracer molecules are moved to the next positions by a distribution which is dictated by the diffusion coefficient. A time-series is generated by summing the fluorescence from all tracer molecules in the illumination volume in each time step. The autocorrelation of these fluorescence fluctuations are compared to experimental data. The best fit( least residues) provides the diffusion coefficient.     \label{simulation}}
\end{figure}

The simulations are performed by defining a rectangular volume with top and bottom surface as confining boundaries with reflecting boundary conditions. The lateral sides were considered to have periodic boundary condition. Figure \ref{simulation} explain the simulation box and parameters associated with it. In the simulation volume, around 10 flurophores  were placed randomly and allowed to take a 3D-random walk. The length of each step $\Delta r$  is picked  from a Gaussian distribution  with a diffusion coefficient($D$).

\begin{align}
\Delta r(r,t)&=\frac{1}{\sqrt{6 \pi D t}} exp \left(\frac{-r^2}{6 D t}\right)\label{eq:distribution}
\end{align}

We can compute positions of all flurophores after every time-step. Assuming a certain excitation intensity profile $I(x,y,z)$  and collection efficiency function $CEF(x,y,z)$ and  using the intensity profile and the absorption cross section, we can calculate number of photons absorbed by the fluorophore. The number of emitted photons can be calculated by assuming a typical quantum efficiency.
The  product of the $CEF$ with the number of emitted photons gives the fraction of photons detected from each flurophore in various positions in the box. Summing the detected photons will give the fluorescent intensity of that instant. Thus, after every time step we can compute photon count from all flurophores in the volume and  time series of fluorescent intensity can be recorded. Simulations were repeated with varying diffusion coefficient until the resulting autocorrelation curve matches  with experimental data.

In conventional FCS, an objective is used for both illumination  and collection of photons.The illumination profile is given by \cite{siegmanlasers}. 

\begin{align}
I(r,d_{xy})&=\frac{2 P}{\pi d_{xy}^2} exp \left(\frac{-2r^2}{d_{xy}^2}\right)\label{eq:obj-illu}\\
where,\nonumber \\
r&= \sqrt{x^2+y^2}\nonumber\\
d_{xy}&=\omega_o \sqrt{1+\left(\frac{z \lambda}{\pi \omega_o^2}\right)^2} , \omega_o=\frac{0.61 \lambda}{NA}\nonumber
\end{align}

Here, $P$ and $\lambda$ are the power and wavelength of light and $NA$, the numerical apperture of the objective. This $I(r,d_{xy})$ is used to calculate number of photons emitted from flurophores in different locations of the simulation box. Governed  by  the confocal arrangement, a fraction of these emitted photons are detected. To calculate that fraction, we can define a collection efficiency function $CEF(x,y,z)$ as
  
\begin{align}
CEF(x,y,z)&= exp \left(\frac{-2(x^2+y^2)}{\Omega_{xy}^2}+\frac{-2z^2}{\Omega_z^2}\right) \label{eq:cef}
\end{align}

\begin{figure}
      \includegraphics[width=0.5\textwidth]{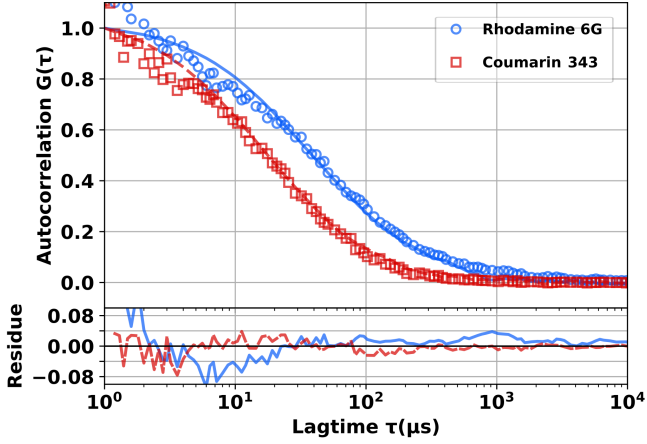}
  \caption{ Conventional FCS measurements using Rhodamine 6G(open circles) and Coumarin 343 (open squares). The continuous  lines are simulated autocorrelations curves by fixing the diffusion coefficient of these dyes to a value determined from previous bulk measurements. The calibration allows us to determine the parameters $\Omega_{xy}$ and $\Omega_z$ to be used in the simulations. $\Omega_{xy}$ = 486 nm  for  532 nm excitation used for Rhodamine 6G  and  $\Omega_{xy}$ = 340 nm  for 446 nm excitation used for  Coumarin 343. $\Omega_z$ is 7.8 $\mu$m }
  \label{plot_cali}
\end{figure}

Where $\Omega_{xy}$ and $\Omega_{z}$ are the Gaussian radius along xy-plane and z-direction where the collection efficiency falls to $1/e^2$. These parameters depend on the size of the pinhole and wavelength of the light. For determining diffusion coefficient $D$ by comparing simulations to experimental data we need to determine $\Omega_{xy}$ and $\Omega_z$  for pinholes used in our experiments.  The diffusion coefficient $D$  of Coumarin 343 in bulk water is known to be 550 $nm^2/\mu$s\cite{c343}. We fixed it in simulation runs and varied $\Omega_{xy}$ and $\Omega_{z}$  until it matches the experimental data obtained for bulk experiments. We obtained $\Omega_{xy}$ = 340 nm and $\Omega_z = 7.8  \mu$m for 446 nm excitation. Similarly, for Rhodamine 6G with $D$ = 400 $nm^2/\mu$s\cite{rhod6g}, we obtained $\Omega_{xy}$ = 486 nm and $\Omega_z = 7.8  \mu$m for 532 nm excitation. See figure \ref{plot_cali}.

In case where tip illumination is used,the illumination profile differs from conventional FCS.  To describe it, the excitation intensity from the tip is approximated as a Gaussian along xy-plane with increasing width $d_{xy}$ away from the tip, determined by a half-cone angle $\alpha$. This intensity is normalized to have constant power($P$) in xy-plane at any z-position. The half-cone angle($\alpha$) can vary from $5.6^o$ (NA of the fiber used) to approximately $30^o$ (for the tip). At a distance of few nm away from the tip aperture, $d_{xy}$ does not change appreciably from radius of tip aperture $r_{tip}$ for these values $\alpha$. 
The illumination profile then becomes
\begin{align}
I(r,d_{xy})&=\frac{2 P}{\pi d_{xy}^2} exp \left(\frac{-2r^2}{d_{xy}^2}\right)\label{eq:tip-illu}\\
where,\nonumber\\
r&= \sqrt{x^2+y^2} \nonumber\\
d_{xy}&=r_{tip}+tan\left(\alpha \right)|z|\nonumber
\end{align}
 The collection efficiency is given by equation \ref{eq:cef}

In this scenario, we need to use the tip radius $r_{tip}$ in the simulation which is determined from size of optical aperture observed in the SEM pictures of the tip. We used $\Omega_{xy}$ and $\Omega_{z}$ determined from the conventional FCS experiments.  To obtain diffusion coefficient $D$ under confinement, we fixed $\Omega_{xy}$, $\Omega_{z}$ and  $r_{tip}$, and performed simulation runs by varying $D$ until the simulation data matches with experiments.

\begin{figure}
      \includegraphics[width=0.5\textwidth]{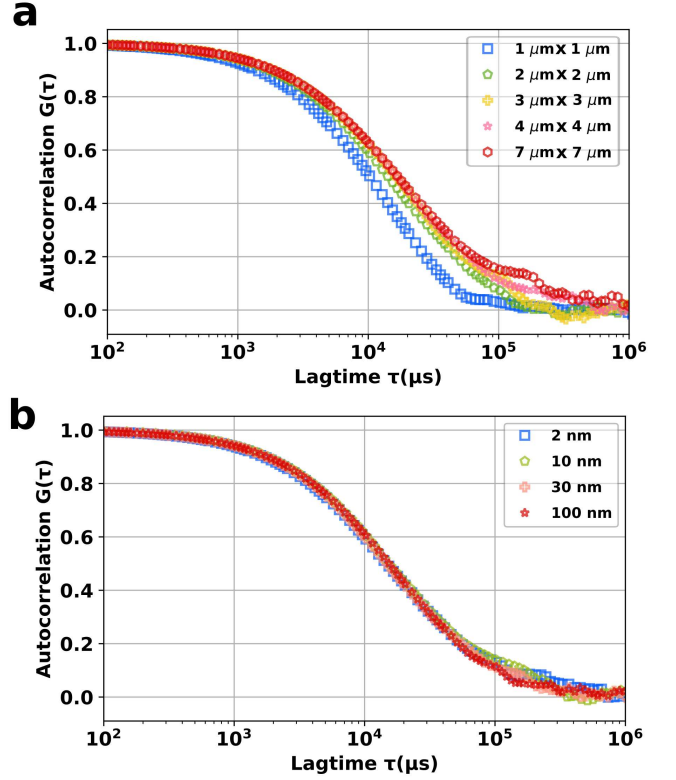}
  \caption{a) The  simulated autocorrelation curves overlap each other if lateral area of the  simulation box is larger that 3 $\mu$m $\times$ 3 $ \mu$m.  We choose the simulation box size to be larger than 3 $\mu$m $\times$ 3 $ \mu$m,The height of the simulation box is kept fixed at 2 nm. b) the height to simulation box does not affect the autocorrelations if it is smaller than $\Omega_z$. the lateral area is fixed at 3 $\mu$m $ \times$ 3 $ \mu$m. The diffusion coefficient used here is 1.3 $nm^2/\mu$s  } 
  \label{plot_box}
\end{figure}

For simulating our experiments, we used 3 $\mu$m $\times$ 3 $\mu$m simulation box with height of few nm.  We found that it is important to take  lateral length which is 3 times or larger than the tip aperture-diameter. It ensures that  the simulated autocorrelation curve is  independent of the lateral-size of the simulation box and avoids artefacts owing to the periodic boundary conditions used. See figure \ref{plot_box} a. Similarly, the simulated autocorrelation should not depend on the vertical dimension of simulation box as long as it is smaller than $\Omega_z$. See figure \ref{plot_box} b.  We used 1 $\mu $M as the fluorophore concentration, which corresponds to 11 flurophores in the simulation box. Increasing the concentration  decreases the intercept to the ordinate of the simulated autocorrelation curve. The normalized autocorrelations for different concentrations overlapped for given diffusion coefficient(data not shown). The simulations generated fluorescent intensity time trace for 5 s with 5 $\mu$s time step. 
We performed few control runs to ensure the validity of our simulations. When we increased the separation between top and bottom boundary of the simulation box ( $\geq$ 3 $\mu$m ) and used collection efficiency function with $\Omega_{xy} = 340$ nm, the bulk measurements match the simulation data for the reported value of bulk diffusion coefficient of Coumarin 343 (550 $nm^2/\mu$s). A similar method is used by  Wohland et al. \cite{wohland} to quantify errors in estimating $D$ in the context of conventional FCS.   

\section{RESULTS  AND DISCUSSION}

\begin{figure}
      \includegraphics[width=0.5\textwidth]{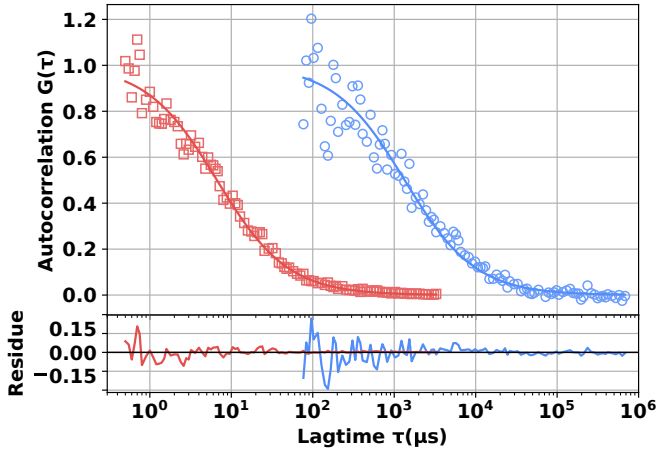}
  \caption{Autocorrelation of fluorescence from Coumarin 343 molecules. The red open squares are experimental data from conventional FCS measurements where  illumination and collection  is performed by the same objective. The continuous red line is fit by equation \ref{eq:2dfcs_wxy}. The procedure allows to extract the parameters such as $\omega_{xy}$  and $\omega_z$. The blue open circles is experimental autocorrelation of fluorescence intensities from the tracer molecules confined between the tip and the substrate for a separation of less than 5 nm. The continuous blue line is fit by equation \ref{eq:2dfcs_wxy} }
  \label{plot1}
\end{figure}

\begin{figure}
      \includegraphics[width=0.5\textwidth]{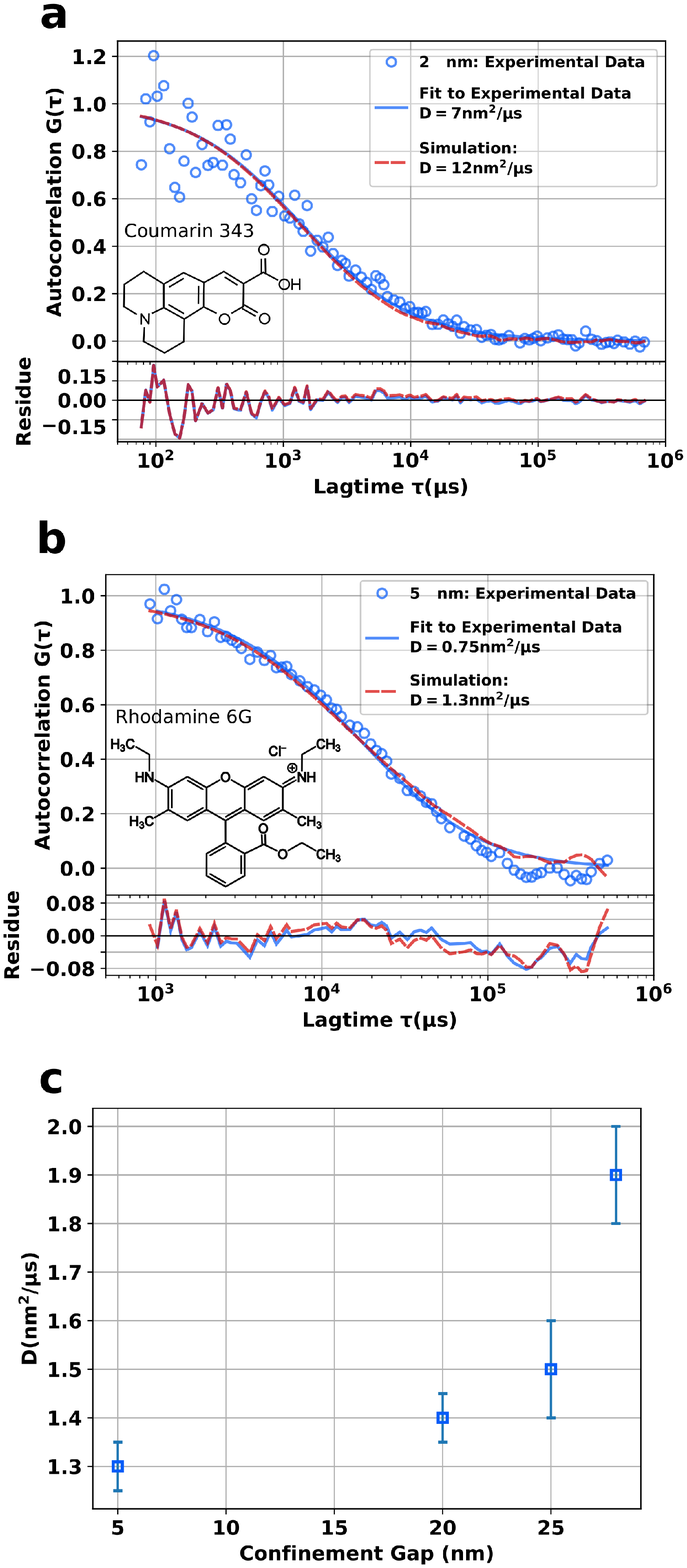}
      
        \caption{ (a)The plot of experimental autocorrelation curves of Coumarin 343(Open circles) collected from confined water($\sim 2$nm), the fit of equation \ref{eq:2dfcs_wxy} with $\omega_{xy}$ = 193 nm (continuous line) and the simulation data(dashed line). (b) The auotcorrelation curves of Rhodamine 6G(open circles),the fit of equation \ref{eq:2dfcs_wxy} with $\omega_{xy}$ = 270 nm (continuous line) and the simulation data(dashed line).    In simulation runs, the parameters are fixed to values determined from the experiments and  the diffusion coefficient is varied till simulated autocorrelation matches the experimental data. (c) The diffusion coefficient estimated by fitting simulated autocorrelation curves  to experimental data on Rhodamine 6G at four different confinement gaps. 
        \label{plot2}}
\end{figure}

 We first fit equation \ref{eq:2dfcs_wxy}  to  experimental data in a figure \ref{plot1}. Similarly, we fit  simulations to experiments as shown in figure \ref{plot2}. We compare the resulting fit parameters.

Figure \ref{plot1} shows the auto-correlation of fluorescent fluctuations from freely diffusing Coumarin 343 in the bulk and from confined volume. In bulk water, it's  diffusion coefficient is known to be $D$= 550 $nm^2/{\mu}$s \cite{c343}. We use it in the equation \ref{eq:2dfcs_wxy} and determine the parameter $\omega_{xy}$. This $\omega_{xy}$ is used when we fit equation \ref{eq:2dfcs_wxy} to the autocorrelation of fluorescence from the experiments wherein the tracer molecule diffuses in confined water. The procedure allows to extract diffusion coefficient of Coumarin 343 in water under confinement between two surfaces separated by 2 nm. We find that the diffusion coefficient of 7.1 $nm^2 / {\mu }$s  is obtained from  fitting \ref{eq:2dfcs_wxy} to our data.  It  suggests that diffusion is slowed down by a factor of ~80 compared to the bulk value(550 $nm^2 / {\mu}$s). 

The diffusion coefficient under confinement determined in this way could be riddled with errors. This is because the equation is derived for conventional FCS, whereas in our experiments the fluorophore is diffusing in a volume bounded by physical boundaries along z- direction. Equation \ref{eq:3dfcs_wxy} or \ref{eq:2dfcs_wxy}  is valid for illumination profile when experiment is performed with an objective and not by fiber-tip. 

As discussed in section IV, in-order to estimate the diffusion coefficient accurately we developed a Monte-Carlo  method which can accommodate our  experimental parameters. In figure \ref{plot2} we plot the simulated auto-correlation curves which fits best to the experimental data on Coumarin 343 and Rhodamine 6G. The simulation runs are performed by varying the diffusion coefficient till it fits the experiments.  We find that this procedure gives a diffusion coefficient of 12 $nm^2/ \mu$s. Thus, the diffusion coefficient in water confined below 2 nm is seen to be 40 times smaller than the bulk coefficient.Similarly, we obtained the diffusion coefficient of Rhodamine 6G  below 2 $nm^2/\mu$s under confinement. It is $\sim$ 400 times smaller than  bulk diffusion coefficient of Rhodamine 6G in water. Figure \ref{plot2} c shows diffusion coefficient at four different tip-substrate separations for Rhodamine 6G. For Coumarin 343, no discernible variation in diffusion coefficient is seen with respect to separation below 20 nm. It is known that Rhodamine 6G possesses positive charge. The enhanced reduction in diffusion  and its variation with respect to separation is possibly due to electrostatic interaction with the negative charge on the glass accumulated during UV treatment. A large slow-down at separations as large as 25 nm  is possibly the effect of partially surface bound flurophores due to this interaction.  More experiments are required wherein,  surfaces are carefully engineered for varying wettability, roughness and charge, before a clear conclusion about tracer diffusion under confinement is drawn. The data presented here is preliminary and asserts the capability of quantitative estimation of  diffusion under confinement using our instrument. The error bars in figure \ref{plot2} c are estimated from residuals after the fit procedure. The simulated autocorrelation curves which have similar residues  after fitting to experimental data decide the upper and lower bound on diffusion coefficient

The estimated diffusion coefficient is likely to be an underestimate due to a possible overestimate of the observation volume used in the simulations. The parameters used in the simulation such as, $\Omega_{xy}$, $\Omega_z$,  tip aperture radius($r_{tip}$) and the confinement gap are determined from the experiments.  Since these parameters describe the observation volume,  a possible overestimate of observation volume  may arise from  errors in their measurement. In simulations, the observation volume is the result of  the convolution of collection efficiency function  and the illumination function. Illumination function is mainly determined by tip aperture radius $r_{tip}$.  The $r_{tip}$  ($\sim$ 400 nm ) used for illumination is bigger than $\Omega_{xy}$ ($\sim$ 340 nm). The confinement gap or the tip-substrate distance ($\sim$ 2nm) is much smaller than the height of the $\Omega_z$. Hence, the observation volume is mainly determined by tip-substrate distance and the $\Omega_{xy}$. As expected, we found that increasing the tip-aperture $r_{tip}$ or the $\Omega_z$ in the simulations does not affect the estimation of diffusion coefficient.  

In order to obtain the upper bound on $D$, we fix $\Omega_{xy}$ to a value which is 50\% more than experimentally measured, the diffusion coefficient increases to  D=18 $nm^2/\mu$s. This is still 30 times smaller than bulk diffusion coefficient of Coumarin 343 in bulk water. Another parameter which may lead to an incorrect estimate of the 
 observation volume is confinement gap or the height of the simulation box.We observed that increasing the gap size by 50 \% does not affect the estimate of diffusion coefficient.

 There are reports of extending the range of FCS application to higher concentrations and its realm to measure  diffusion in biological membranes.\cite{FCSwithNSOM1,FCSwithNSOM2,FCSwithNSOM3,FCSwithNSOM_membranes1,FCSwithNSOM_membranes2}.In these works, sub-diffraction-limited illumination is used to perform measurements in bulk and 2-D membranes. In our instrument  the illumination is varied from 150 nm to  1 $\mu$m which results in better throughput of data.These methods lack  special data analysis methods to take into account experimental geometry used for confinement and its physical effects.  
 
Zhong et al.\cite{Zhong} used sub- 10 nanometer channels to image diffusion of Rhodamine B using optical microscope. This has been made possible by a silver nitride layer below the channels to form a Fabry-Perot interference which enhances the fluorescence from the tracer dye and it is detectable under optical microscope. It allows them to measure the concentration gradients with respect to time. After fitting a convection-diffusion model to this data, they estimated diffusion coefficient of Rhodamine B to be  2.1 $nm^2/\mu$s,  a slow-down by about 100 times compared to bulk. While the preparation of channels is an involved task, the advantage of this method lies in simplicity of measurement and analysis. In our method we employ a feedback mechanism to maintain the confinement gap below 10 nm. While this is cumbersome while performing measurements, our method allows changing the gap size (channel depth) continuously. It also facilitates replacement of confining substrate  with relative ease. In FCS methodology equilibrium fluctuations are used to estimate diffusion coefficient whereas Zhong et al.\cite{Zhong} have measured concentration gradients to estimate diffusion coefficient. It is noteworthy that both these methods are giving similar diffusion coefficients ($\sim$ 1.3 $nm^2/\mu$s and 2.1 $nm^2/\mu$s)  for  sub- 10 nm confinement.

Santo et al.\cite{Santo} have used FCS combined with nanochannels to measure diffusion coefficient of confined polymers as well as Rhodamine 6G. They  observed subdiffusive behavior for polymers confined to nanochannels. However, diffusion of Rhodamine 6G   in 10 nm channels was not significantly altered compared to bulk. This is not in agreement with  our observation as well as that of Zhong et al.   The possible reason behind this could be overnight equilibration  of  channels with the solvent. This may have reduced  the  possible electrostatic interactions between the dye and channel surface. Since both the illumination and collection is performed using same objective, a conventional FCS theory is used to fit experimental data\cite{Santo}.  However, in our method we have used tip illumination and its profile needs to be incorporated into the analysis. As shown in figure \ref{plot2}a and b, this  discrepancy in estimating diffusion coefficient is clearly seen.

It is  known that Coumarin 343 has negative charge. The cover-slips used in our experiments are cleaned with  UV treatment which also renders them negatively charged. We attribute a  much larger slow-down ( $\sim$ 300 times) of  Rhodamine 6G compared to Coumarin 343 ($\sim$ 50) to the surface charge on glass. It is noteworthy that the Coumarin 343 diffusion coefficient is reduced by  roughly 50 times compared to bulk even though the surfaces and the dye are both negatively charged. This points to a  need for more careful preparation of surfaces  before we conclude about the tracer diffusion under nanoconfinement.    

There are other attempts to study diffusion in confined liquids\cite{Ashish,Bocquet}by combining FCS with other techniques. Mukhopadhyay et al.\cite{Ashish} measured  autocorrelation of fluorescence from the tracer molecules confined in  molecularly thin films of liquid in Surface Force Apparatus(SFA). In this experiment, the liquid is confined between two mica sheets wrapped around crossed cylinders.  They used interference to deduce the separations with special optical windows on the mica to carry out the experiments. This amounts to  applying different coatings for the optical window for different excitation wavelengths. In our instrument, we can change the laser coupled  to the fiber for different flurophores with relative ease. Joly et al.\cite{Bocquet} used FCS to measure diffusion of colloidal fluorescent microspheres diffusing in a gap between a cover-slip and convex lens placed on it. The confinement-gap can be changed by shifting the confocal observation volume to different radial positions. In this study they used submicron sized fluorescent-beads instead of molecular probes which limits the confinement size to the bead size $\geq 100$ nm.  In this experiment, the gaps of the order of few nm are difficult to achieve. 

Our instrument can  be used to measure tracer diffusion in confined liquids besides water, in particular of liquid electrolytes or lubricants. It can also be  used for experimental testing of recent calculations to describe the diffusion in confined liquids with rough or patterned confining walls\cite{entropicbarrier_pre,entropicbarrier_bocquet}.

In conclusion, we  have developed a novel instrument to measure diffusion in water confined to 1-10 nm separation. A Monte-Carlo method is developed to quantify measured  quantities to obtain diffusion coefficient in such nanoscale confinement. Together, it provides a unique and versatile method to measure diffusion in nanoconfined water.   

\section{ACKNOWLEDGEMENTS}
Authors would like to thank Dr. Thorsten Wohland from National University of Singapore for his useful comments. 
VJ Ajith would like to acknowledge the INSPIRE PhD fellowship from Department of Science and Technology(DST), Govt. of India. The work is partially funded by Wellcome-Trust DBT India Alliance. A portion of this research was performed using facilities at CeNSE, Indian Institute of Science, Bangaluru, funded by the Ministry of Electronics and Information Technology(MeitY), Govt. of India.



\bibliography{Ref}{}
\bibliographystyle{aipnum4-1}

\end{document}